\documentclass[pra,showpacs,twocolumn,superscriptaddress]{revtex4-1}
\usepackage{mathrsfs}
\usepackage{amsfonts}
\usepackage{amsmath}
\usepackage{txfonts}
\usepackage{amssymb}
\usepackage{graphicx,subfigure}
\usepackage{bm}
\usepackage{color}
\usepackage[normalem]{ulem}

\newcommand{\ket}[1]{|#1\rangle}
\newcommand{\bra}[1]{\langle #1|}

\begin{document}
\title{Genuinely noncyclic geometric gates in two-pulse schemes}
\author{Nils Eivarsson}
\affiliation{Department of Physics and Astronomy, Uppsala University,
Box 516, Se-751 20 Uppsala, Sweden}
\author{Erik Sj\"oqvist}
\email{erik.sjoqvist@physics.uu.se}
\affiliation{Department of Physics and Astronomy, Uppsala University,
Box 516, Se-751 20 Uppsala, Sweden}
\date{\today}
\begin{abstract}
While most approaches to geometric quantum computation are based on 
geometric phase in cyclic evolution, noncyclic geometric gates have 
been proposed to further increase flexibility. While these gates 
remove the dynamical phase of the computational basis, they do not 
in general remove it from the eigenstates of the time evolution 
operator, which makes the geometric nature of the gates ambiguous. 
Here, we resolve this ambiguity by proposing a scheme for 
{\it genuinely noncyclic geometric gates}. These gates are obtained 
by evolving the computational basis along open paths consisting 
of geodesic segments, and simultaneously assuring that no dynamical 
phase is acquired by the eigenstates of the time evolution operator. 
While we illustrate the scheme for the simplest nontrivial case of 
two geodesic segments starting at each computational basis state of 
a single qubit, the scheme can be straightforwardly extended to more 
elaborate paths, more qubits, or even qudits. 
\end{abstract}
\maketitle
\date{\today}

\section{Introduction}
Quantum computation is a form of information processing that uses 
quantum mechanical properties, such as superposition and entanglement, 
to perform calculations. Although quantum computers of today do not 
have the number of qubits required for realising programmable 
large-scale computation, progress towards this goal has been 
achieved \cite{debnath16} on different experimental platforms 
\cite{watson18,wu19,roy20}. 

It is not, however, enough to scale up quantum computers; they also 
need to be resilient to noise and decoherence under which qubits 
would lose their desired quantum mechanical properties. The state 
within a quantum computer cannot be completely isolated and decoherence 
is therefore inevitable. It is possible to overcome this problem 
according to the threshold theorem \cite{nielsen00,gottesman10}, 
which entails that an arbitrarily long quantum computation 
can be sustained by error correction techniques provided the error 
rate per gate is below a certain threshold value. 

One approach to reach below this threshold is to use geometric 
quantum gates \cite{jones00,ekert00,xiang01,zhu02,zhu03,wang07,liu19}. 
These are based on the 
(Abelian) geometric phase (GP) of quantum systems 
\cite{berry84,aharonov87}. When a quantum state undergoes a 
cyclic evolution it can gain a phase factor. This phase factor 
can be split into a dynamical part and a geometric part, where 
the geometric part is only dependent on the path a quantum state 
takes through its state space. By choosing an evolution for which 
the dynamical phase effect is trivial, it is possible to implement 
quantum logic gates that are purely dependent on the geometry 
of the path. This dependency on only geometry can be shown to 
be resilient against certain types of noise 
\cite{leibfried03,zhu05,ota09,kondo11,ichikawa12,berger13,ding22,liang22}. 

While geometric quantum gates show promise in the implementation 
of robust quantum computers, there are still improvements to be 
made before they are fully operational. One possible improvement 
is to reduce the number of pulses and the run time needed 
to implement the gates, as this would reduce the qubits' 
exposure to the environment,  
thereby limiting the effect of  decoherence, as well as simplifying the pulse 
schemes. To reduce the run time, noncyclic evolution in the 
implementation of geometric quantum gates have been proposed 
\cite{friedenauer03,ericsson08,wang09,liu20,ji21,qiu21,sun22,guo22} 
and experimentally implemented \cite{zhang21}. These gate are based 
on computational states evolving along open paths while still only 
acquiring a GP in noncyclic evolution \cite{samuel88,mukunda93}. 
However, while the GP factors acquired by the computational states are 
eigenvalues of the time evolution operator when implementing cyclic geometric 
gates, this is not so in noncyclic evolution since in this case the eigenstates 
do not coincide with the computational states. This means that the eigenstates 
may pick up additional nontrivial dynamical phases also in cases where the 
noncyclic gates look geometric in the computational basis. Thus, the geometric 
meaning of these noncyclic gates is ambiguous. 

Here, we propose precise conditions for geometric gates. Under 
these conditions the meaning of noncyclic geometric 
gates become unambiguous. This provides a notion of {\it genuinely 
noncyclic geometric gates}. Furthermore, geometric gates under 
cyclic evolution of the computational states, which are gates 
that trivially satisfy these conditions, are Abelian (they 
diagonalize in the computational basis) and therefore fail 
universality. In order to complete the universal set, we shall 
thus see that at least one genuinely noncyclic geometric gate 
is needed. 

\section{Conditions for geometric gates} 
We consider Schr\"odinger evolution  
$i\hbar \ket{\dot{\psi} (t)} = H(t) \ket{\psi (t)}$, 
where $H(t)$ is the Hamiltonian of the system with Hilbert 
space $\mathcal{H}$. Let $U(t,0) = {\bf T} 
e^{-\frac{i}{\hbar} \int_0^t H(t') dt'}$ be 
the corresponding time evolution operator. We 
assume that the computational system consists of $n$ 
qubits with state space $\mathcal{M} \subseteq \mathcal{H}$ 
spanned by $2^n \leq \dim \mathcal{H}$ predetermined computational 
state vectors $\{ \ket{\vec{q}} = \ket{q_1,\ldots,q_n} 
\}_{q_1,\ldots,q_n=0,1}$, fixed by the final read out of the computation. 

Let $U(\tau,0)$ be the desired gate realized during the time 
interval $t\in [0,\tau]$. Consider the eigenvalue equation 
\begin{eqnarray}
U(\tau,0) \ket{\psi_k} = e^{i\varphi_k} \ket{\psi_k}, \ k=1,\ldots,2^n.
\end{eqnarray}
Here, $\{ \ket{\psi_k} \}$ is an orthonormal set of vectors since 
$U(\tau,0)$ is a normal operator \cite{nielsen00}. We  
assume that $U(\tau,0)$ preserves $\mathcal{M}$, i.e., 
$\ket{\psi_k} = \sum_{\vec{q}} c_{\vec{q}}^{(k)} \ket{\vec{q}}$, 
$\forall k$, even in the case where $2^n < \dim \mathcal{H}$. 
Thus, $U(\tau,0) U^{\dagger} (\tau,0)$ is a projection operator on $\mathcal{M}$. 

The action of $U(\tau,0)$ on each computational state $\ket{\vec{q}}$ 
can be understood in terms of its GP  
\begin{eqnarray}
\Gamma_{\vec{q}} & = & \arg \bra{\vec{q}} U(\tau,0) \ket{\vec{q}} 
\nonumber \\ 
 & & + i\int_0^{\tau} \bra{\vec{q}}
U^{\dagger} (t,0) \dot{U} (t,0)
\ket{\vec{q}} dt , 
\label{eq:general_gp}
\end{eqnarray}
provided $\bra{\vec{q}} U(\tau,0) \ket{\vec{q}}$ is nonzero. 
$\Gamma_{\vec{q}}$ is real-valued and a property of the path 
$U(t,0) \ket{\vec{q}}\bra{\vec{q}}U^{\dagger} (t,0)$ in state 
space, as it is invariant under monotonic reparametrizations 
$t\mapsto s(t)$ and  time-local phase changes $U(t,0) \mapsto 
e^{if (t)} U(t,0)$ \cite{mukunda93}. These $\Gamma_{\vec{q}}$:s 
are generically noncyclic GPs as the paths in state space are 
typically open. On the other hand, the eigenvectors $\ket{\psi_k}$ 
of $U(\tau,0)$ are the cyclic states of the evolution $U(t,0)$, 
$t\in [0,\tau]$, and may thus be analyzed by using the 
Aharonov-Anandan GP \cite{aharonov87}. In this framework, 
each cyclic phase 
$\varphi_k$ contains a geometric ($\gamma_k$) and a dynamical 
($\delta_k$) part, given by 
\begin{eqnarray}
\gamma_k & = & \arg \bra{\psi_k} U(\tau,0) \ket{\psi_k} 
\nonumber \\ 
 & & + i\int_0^{\tau} \bra{\psi_k} U^{\dagger} (t,0) 
\dot{U} (t,0) \ket{\psi_k} dt . 
\label{eq:general_gp}
\end{eqnarray} 
and 
\begin{eqnarray}
\delta_k = - \frac{1}{\hbar} \int_0^{\tau} \bra{\psi_k} 
U^{\dagger} (t,0) H(t) U (t,0) \ket{\psi_k} dt , 
\label{eq:general_dp}
\end{eqnarray}
such that $\varphi_k = \gamma_k + \delta_k$.

Based on the above, we can now define a {\it genuinely 
noncyclic geometric gate} (GNGG) as a unitary $U(\tau,0)$ that 
satisfies the following two conditions: 
\begin{itemize}
\item[(i)] $\Gamma_{\vec{q}} - \Gamma_{\vec{0}} = 
\arg \bra{\vec{q}} U(\tau,0) \ket{\vec{q}} - \arg \bra{\vec{0}} 
U(\tau,0) \ket{\vec{0}}, \! \! \mod 2\pi$,
\item[(ii)] $\delta_k - \delta_1 = 0, \! \! \mod 2\pi$,
\end{itemize}
for binary vectors $\vec{q}$ and $k$. In essence, such 
a gate depends only on GPs for both its eigenstates 
and the predetermined computational states. 

Before addressing the physical realization of GNGGs in the next 
section, we consider some qubit gates assuming 
them to be genuinely noncyclic, in order 
to gain some further conceptual insights. First, let us consider a 
genuinely noncyclic geometric Hadamard gate $\mathbb{H}$. 
Such a gate takes the form 
\begin{eqnarray}
\mathbb{H} = \frac{1}{\sqrt{2}} \Big( \ket{0} + \ket{1} \Big) \bra{0} 
+ e^{i\pi} \frac{1}{\sqrt{2}} \Big( \ket{1} - \ket{0} \Big) \bra{1}  
\end{eqnarray}
in the computational basis \cite{remark1,pancharatnam56}, and 
\begin{eqnarray}
\mathbb{H} & = & \ket{+} \bra{+} + 
e^{i\pi} \ket{-} \bra{-} 
\end{eqnarray}
in the eigenbasis $\ket{\pm} = \frac{1}{4 \mp 2\sqrt{2}} \left[ \ket{0} - 
\left( 1 \mp \sqrt{2}\right) \ket{1} \right]$, with the phase factor $e^{i\pi} = -1$ 
assumed to be geometric in both cases. We thus see that the geometric phase 
difference for both forms is $\pi$, i.e., $\Gamma_1-\Gamma_0 = 
\gamma_2-\gamma_1 = \pi$.  

As a second example, we consider rotation gates 
$\mathbb{U}_z (\vartheta_z) = 
e^{-i\frac{\vartheta_z}{2} \sigma_z}$ and 
$\mathbb{U}_y (\vartheta_y) = 
e^{-i\frac{\vartheta_y}{2} \sigma_y}$ that can 
be used to describe an arbitrary single-qubit 
rotation \cite{sakurai93}. First, we note that 
$\mathbb{U}_z (\vartheta_z)$ is diagonal in the 
computational basis, i.e., the eigenbasis 
coincides with the computational basis. This 
implies that the gate is a standard cyclic 
geometric gate \cite{ekert00}, for which 
$\Gamma_1 - \Gamma_0 = \gamma_2 - \gamma_1 = 
\vartheta_z$. More interesting is the genuinely 
noncyclic geometric implementation of 
$\mathbb{U}_y (\vartheta_y)$. We may write
\begin{eqnarray}
\mathbb{U}_y & = & \left( \cos \frac{\vartheta_y}{2} \ket{0} + 
\sin \frac{\vartheta_y}{2} \ket{1} \right) \bra{0} 
\nonumber \\ 
 & & + \left( -\sin \frac{\vartheta_y}{2} \ket{0} + \cos 
\frac{\vartheta_y}{2} \ket{1} \right) \bra{1}
\end{eqnarray}
and 
\begin{eqnarray}
\mathbb{U}_y & = & 
e^{-i\frac{\vartheta_y}{2}} \ket{y_+} \bra{y_+} + 
e^{i\frac{\vartheta_y}{2}} \ket{y_-} \bra{y_-}, 
\nonumber \\ 
\ket{y_{\pm}} & = & \frac{1}{\sqrt{2}} \big( \ket{0} 
\pm i\ket{1} \big)
\end{eqnarray}
in the computational basis and eigenbasis, respectively. 
Provided $\vartheta_y \neq \pi$, we find $\Gamma_1 - 
\Gamma_0 = 0$ and $\gamma_2 - \gamma_1 = \vartheta_y$, 
i.e., while the noncyclic GP difference of the computational 
basis states is trivial, the cyclic GP difference is 
generally not. In the case where 
$\vartheta_y = \pi$, the noncyclic phases are not 
defined as the computational states are each mapped on 
orthogonal states. It follows, more generally, that $\pi$ 
rotations around any axis in the $xy$ plane cannot be 
implemented in a genuinly noncyclic manner. Curiously, 
this implies that there is no genuinely noncyclic 
implementation of CNOT, while the GNGG scheme can be 
used to realize other entangling two-qubit control 
gates with noncyclic and cyclic GPs 
$0,0,\Gamma_{10},\Gamma_{11}$ and $0,0,\gamma_3,\gamma_4$, 
respectively. Similarly, the  SWAP gate cannot be implemented as a 
GNGG since it interchanges the computational states $\ket{01}$ and $\ket{10}$, 
which are orthogonal in both qubits, while a genuinely noncyclic realization of 
the $\sqrt{\rm SWAP}$ gate exists. The latter takes the form 
\begin{eqnarray}
\sqrt{{\rm SWAP}} & = & \ket{00} \bra{00} + e^{i\frac{\pi}{4}} \frac{1}{\sqrt{2}} 
\left(  \ket{01} -i \ket{10} \right) \bra{01} 
\nonumber \\ 
 & & + e^{i\frac{\pi}{4}} \frac{1}{\sqrt{2}} \left( -i \ket{01} + \ket{10} \right) \bra{10} + \ket{11} \bra{11} 
\end{eqnarray}
in the computational basis, and \begin{eqnarray}
\sqrt{{\rm SWAP}} & = & \ket{e_1} \bra{e_1} + \ket{e_2} \bra{e_2} + e^{i\frac{\pi}{2}} \ket{e_3} \bra{e_3} 
\nonumber \\ 
 & & + \ket{e_4} \bra{e_4} 
\end{eqnarray}
in the eigenbasis $\ket{e_1} = \ket{00}, \ket{e_2} = \frac{1}{\sqrt{2}} (\ket{01}+\ket{10}), 
\ket{e_3} = \frac{1}{\sqrt{2}} (\ket{01}-\ket{10}), \ket{e_4} = \ket{11}$. One finds the 
noncyclic and cyclic GPs $\Gamma_{00} = \Gamma_{11} = 0, \Gamma_{01} = 
\Gamma_{10} = \frac{\pi}{4}$ and $\gamma_1 = \gamma_2 = \gamma_4 = 0,
\gamma_3 = \frac{\pi}{2}$, respectively. Both these sets define nontrivial GP differences.

\section{Two-pulse single-qubit gates} 
A key point of the GNGG technique is that 
it can be used to reduce the number of pulses to implement a universal set of 
geometric gates. To make this point explicit, we shall now 
examine the physical realization of arbitrary genuinely noncyclic 
single-qubit gates, 
by using the simplest nontrivial case of two pulses. To achieve 
universality with geometric gates in such schemes, adhering 
to the proposed conditions (i) and (ii) above, noncyclic 
geometric gates are required, as cyclic gates are inherently Abelian in that they 
are by definition diagonal in the computational basis and thus insufficient for 
universality \cite{remark2}. 

Let the two pulses be applied during $[0,t_1]$ and 
$[t_1,\tau]$, respectively. By choosing the pulses such 
that they move the computational basis along a pair of 
geodesic segments, the corresponding  dynamical phases 
vanish. After constructing these gates, the dynamical 
phases of the eigenstates of the time evolution operator 
are studied, to find which gates are genuinely geometric, 
i.e., satisfy the condition $2 \delta = 0, \! \! \mod 2 \pi$, 
where we have used that $\delta_0 = -\delta_1 \equiv \delta$. As we shall see, 
this requires a careful tuning of rotation axes and precession angles associated 
with the two pulses. 

\begin{figure}[h!]
\centering
\includegraphics[width=0.3\textwidth]{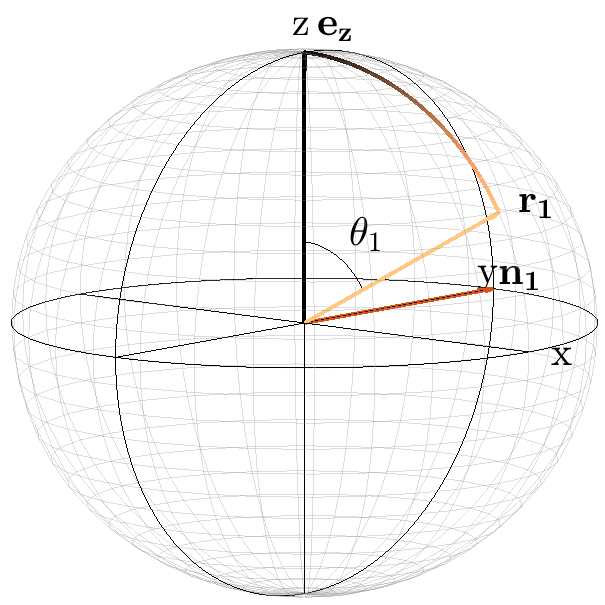}
\includegraphics[width=0.3\textwidth]{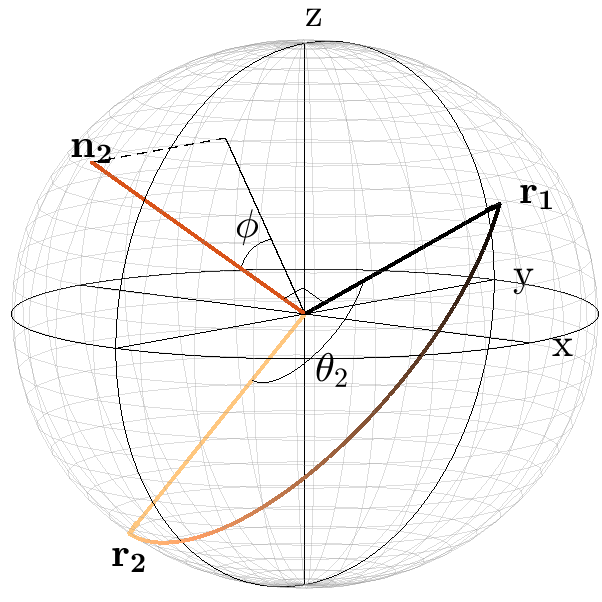}
\caption{The rotation of the qubit during the first pulse (upper panel). 
$\theta_1$ is the rotational angle and ${\bf n}_1$ is the axis 
of rotation. The rotation of the qubit during the second 
pulse (lower panel). $\theta_2$ is the rotational angle and 
$\phi$ determines 
the axis of rotation ${\bf n}_2$. ${\bf n}_2$ is orthogonal 
to the final state (with Bloch vector ${\bf r}_1$) of the 
first pulse. The paths gradually shifts 
from black (dark) at the start to yellow (bright) at the end.}
\label{fig:Rotation}
\end{figure}

A path is geodesic on the Bloch sphere when the axis 
${\bf n}$ of rotation is orthogonal to the initial Bloch 
vector. Since the first pulse acts on the computational basis 
$\{ \ket{q} \}_{q=0,1}$, it should thus correspond to a rotation 
around any axis ${\bf n}_1$ in the $xy$-plane, as this would result 
in evolution of these states along parts of a great 
circle that pass through the poles of the Bloch sphere. Choosing one axis 
over another only changes the eigenvector of the complete evolution by a 
rotation around the $z$-axis. We can therefore limit ourselves to one specific 
axis of rotation and later generalize to other axes in the $xy$-plane by simply 
rotating the gate. We choose the $y$ axis, 
i.e., ${\bf n}_1 = {\bf e}_y$, which defines the 
Hamiltonian 
\begin{eqnarray}
H_1 = \frac{1}{2} \hbar \omega \, \sigma_y 
\end{eqnarray}
with corresponding time evolution operator 
\begin{eqnarray}
U_1(t, 0) = e^{-\frac{i}{2} \omega t} \ket{y_+} \bra{y_+} + 
e^{\frac{i}{2} \omega t} \ket{y_-} \bra{y_-} . 
\label{eq:te_first}
\end{eqnarray}
At the final time $t_1$ of the first pulse, the qubit has rotated 
an angle $\theta_1 = \omega t_1$. This rotation can be seen in the upper 
panel of Fig.~\ref{fig:Rotation}. 

For the second pulse, the axis of rotation must be orthogonal to the final state 
of the first pulse to move 
the qubit state along a geodesic. Starting at $\ket{0}$, the final state 
of the first pulse can be described by the Bloch vector 
${\bf r}_1 = \sin \theta_1 {\bf e}_x + \cos \theta_1 {\bf e}_z$, 
which serves as initial state for the second pulse. The  
axis of rotation can therefore
be taken as 
\begin{eqnarray}
{\bf n}_2 =  -\cos\theta_1 \cos \phi \, {\bf e}_x 
+ \sin \phi \, {\bf e}_y + 
\sin \theta_1 \cos \phi \, {\bf e}_z .
\end{eqnarray}
This lies in the plane spanned by the vector 
$-\cos\theta_1 \, {\bf e}_x + \sin\theta_1 \, {\bf e}_z$, 
orthogonal to ${\bf r}_1$, and the $y$-axis. $\phi$ is 
the rotational angle around ${\bf r}_1$ relative the $xz$-plane, see the lower panel of
Fig.~\ref{fig:Rotation}. With the axis of rotation 
defined, the Hamiltonian becomes
\begin{eqnarray}
H_2 & = & \frac{\hbar \omega}{2} \Big( -\cos\theta_1 
\cos\phi \, \sigma_x + \sin \phi \, \sigma_y 
\nonumber \\ 
 & & + \sin\theta_1 \cos{\phi} \, \sigma_z  \Big) 
\end{eqnarray}
with eigenvalues and eigenvectors 
\begin{eqnarray}
\lambda_\pm & = &  \pm \frac{1}{2} \hbar \omega , 
\nonumber \\
\ket{\kappa_\pm}  & = & \sqrt{\frac{\cos^2 \theta_1 
\cos^2 \phi + \sin^2 \phi}{2 \mp 
2 \sin \theta_1 \cos \phi}} 
\nonumber \\ 
 & & \times \left[\ket{0} + \frac{\mp 1 + 
\sin \theta_1 \cos \phi}{\cos \theta_1 \cos \phi 
+ i \sin \phi} \ket{1} \right].
\end{eqnarray}
The time evolution operator of the second rotation is
\begin{eqnarray}
U_2(t, t_1) =  e^{-\frac{i}{2}  \omega (t-t_1)} \ket{\kappa_{+}} \bra{\kappa_{+}} + 
e^{\frac{i}{2}  \omega \left(t-t_1\right)} \ket{\kappa_{-}} \bra{\kappa_{-}}, 
\label{eq:te_second}
\end{eqnarray}
which rotates the qubit an additional angle $\theta_2 = 
\omega (\tau - t_1)$. This rotation is the geodesic path connecting ${\bf r}_1$ and 
${\bf r}_2$ in the lower panel of Fig.~\ref{fig:Rotation}. The time evolution during the 
full time interval $[0,\tau]$ can be written as 
\begin{eqnarray}
U (t, 0)  = \left\{ \begin{array}{ll}
U_1 (t, 0), & 0 \leq t \leq t_1, \\
U_2 (t, t_1) U_1 (t_1, 0), & t_1 \leq t \leq \tau 
\end{array} \right. 
\label{eq:total_u}
\end{eqnarray}
with the gate $U(\tau,0) = U_2 (\tau, t_1) 
U_1 (t_1, 0)$ being characterized by the angles $\theta_1,\theta_2$, and $\phi$, 
i.e., $U(\tau,0) \equiv \mathbb{U}(\theta_1,\theta_2,\phi)$. 

By inserting Eqs.~\eqref{eq:te_first} and 
\eqref{eq:te_second} into \eqref{eq:total_u}, we 
find the eigenvalues and eigenvectors of $\mathbb{U}(\theta_1,\theta_2,\phi)$: 
\begin{widetext} 
\begin{eqnarray}
\lambda_\pm & = & \cos \left( \frac{\theta_1}{2} \right) \cos \left( \frac{\theta_2}{2} \right) - 
\sin \left( \frac{\theta_1}{2} \right) \sin \left( \frac{\theta_2}{2} \right) \sin  \phi 
\pm i \sqrt{1 -  \left[  \cos \left( \frac{\theta_1}{2} \right) \cos \left( \frac{\theta_2}{2} \right) - 
\sin \left( \frac{\theta_1}{2} \right) \sin \left( \frac{\theta_2}{2} \right) \sin  \phi \right]^2 } , 
\nonumber \\ \
\ket{\psi_\pm} & = & \mathcal{N}_{\pm} \left\{ \ket{0} + 
\frac{ \sin \left( \frac{\theta_1}{2} \right) \sin \left( \frac{\theta_2}{2} \right) \cos  \phi \pm  
\sqrt{1 -  \left[  \cos \left( \frac{\theta_1}{2} \right) \cos \left( \frac{\theta_2}{2} \right) - 
 \sin \left( \frac{\theta_1}{2} \right) \sin \left( \frac{\theta_2}{2} \right) 
 \sin  \phi \right]^2 } } {e^{i \phi} \cos \left( \frac{\theta_1}{2} \right) \sin \left( \frac{\theta_2}{2} 
 \right) + i \sin \left( \frac{\theta_1}{2} \right) \cos \left( \frac{\theta_2}{2} \right)} \ket{1} \right\} 
 \label{eq:eigenvectors}
\end{eqnarray}
with normalization factors 
\begin{eqnarray}
\frac{1}{\mathcal{N}_{\pm}}  = \sqrt{ 1 + \frac{ 2 \left( \sin \left( \frac{\theta_1}{2} \right) 
\sin \left( \frac{\theta_2}{2} \right) \cos  \phi \pm  \sqrt{1 -  \left[  \cos \left( \frac{\theta_1}{2} \right) 
\cos \left( \frac{\theta_2}{2} \right) -  \sin \left( \frac{\theta_1}{2} \right) 
\sin \left( \frac{\theta_2}{2} \right) \sin  \phi \right]^2 } \right)^2}{1 - \cos \theta_1 
\cos \theta_2  + \sin \theta_1 \sin \theta_2 \sin \phi} }
 \label{eq:norm}
\end{eqnarray}
\end{widetext}

\begin{figure}[htp!]
\centering
\includegraphics[width = 0.5\textwidth]{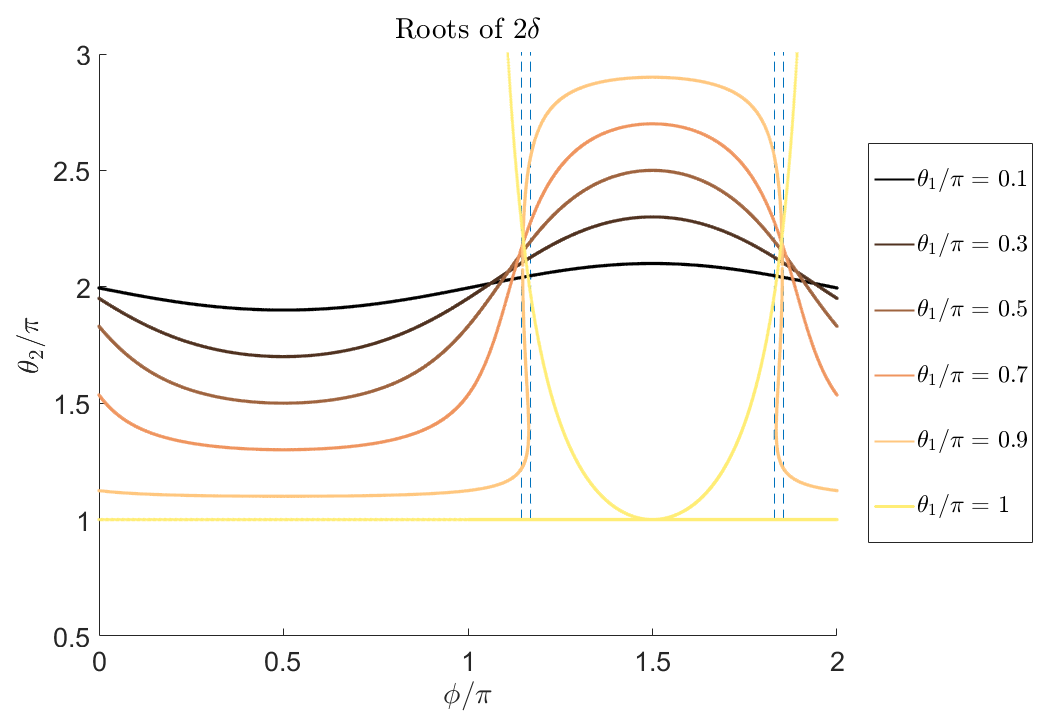}
\includegraphics[width = 0.5\textwidth]{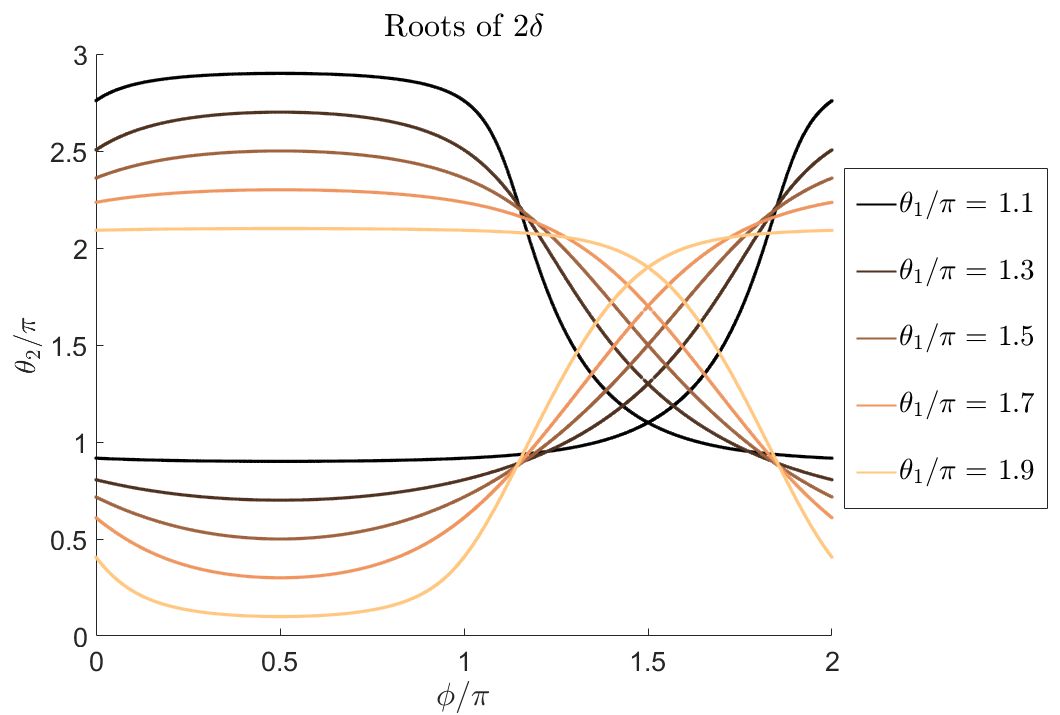}
\caption{The roots of $2 \delta = 0$ (trivial dynamical phase). $\theta_1$ 
($\theta_2$) is the first (second) rotation angle, and 
$\phi$ is the angle determining the axis of rotation 
for the second rotation. In the upper panel, two sets of dashed lines are shown, 
marking where there are three different choices of $\theta_2$ that give zero 
dynamical phase for $\theta_1/\pi = 0.9$. The width of these intervals increases 
when $\theta_1/\pi \rightarrow 1$, until the curve splits into two for $\theta_1=\pi$: 
one concave curve with minimum at $\phi/\pi = 1.5$ and one horizontal line at 
$\theta_2/\pi = 1$, the latter corresponding to the special case of cyclic geometric gates.}
\label{fig:Roots}
\end{figure}

\begin{figure}[htp!]
\centering
\includegraphics[width = 0.5\textwidth]{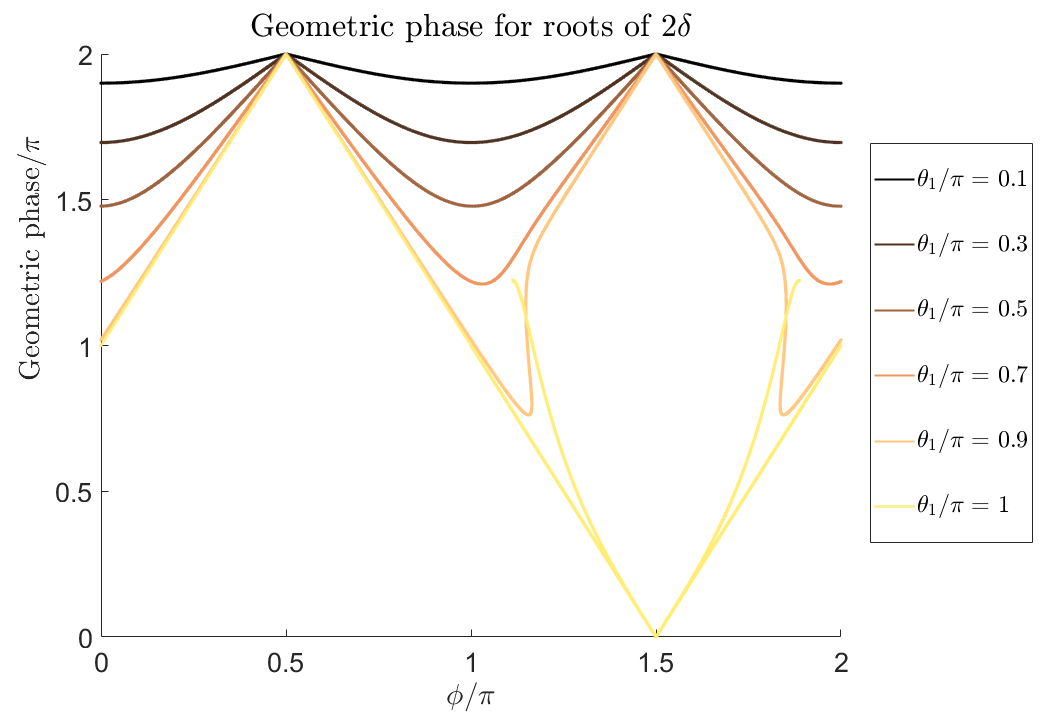}
\includegraphics[width = 0.5\textwidth]{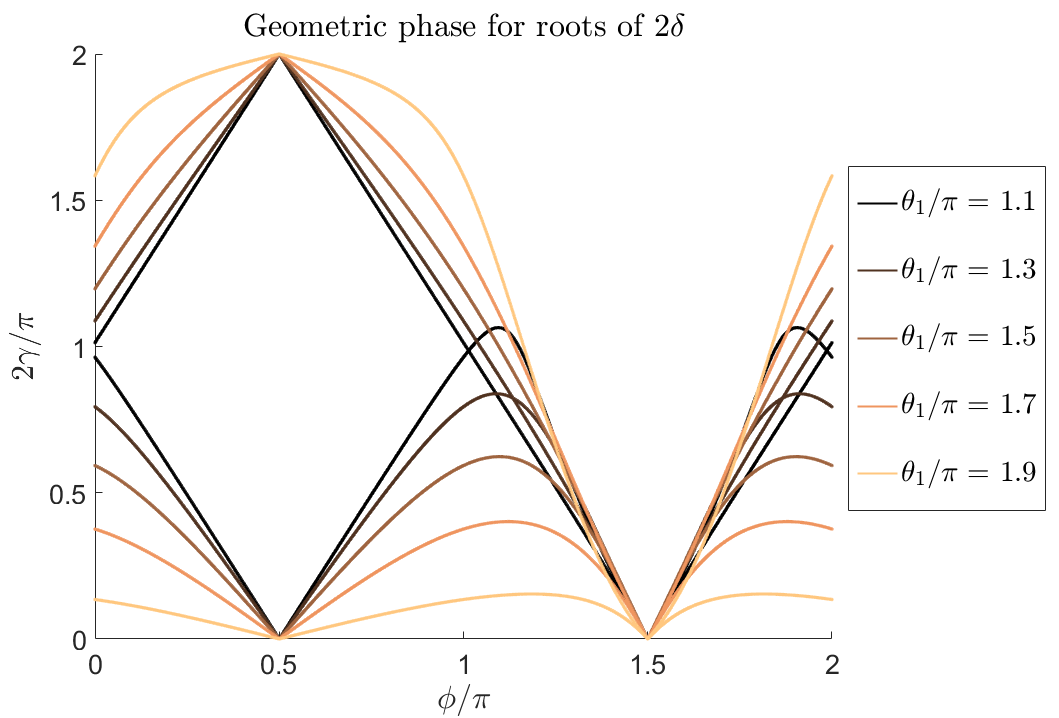}
\caption{GP corresponding to the roots of $2 \delta = 0 $. $\theta_1$ ($\theta_2$) 
is the first (second) rotation 
angle, and $\phi$ is the angle determining the axis of rotation for the second rotation.}
\label{fig:GeometricPhase}
\end{figure}

Next, we find the dynamical phases of the evolution by using 
Eq.~\eqref{eq:general_dp}, yielding 
\begin{eqnarray}
\delta_\pm & = & - \frac{1}{\hbar} \int_0^{t_1} \bra{\psi_\pm} H_1 
\ket{\psi_\pm} dt 
\nonumber \\
 & & - \frac{1}{\hbar}\int_{t_1}^{\tau} \bra{\psi_\pm} U_1^{\dagger}(t_1, 0) 
H_2 U_1(t_1, 0) \ket{\psi_\pm} dt, 
\label{eq:total_te}
\end{eqnarray}
where we have taken into account that $H_1$ and $H_2$ commute with 
$U_1$ and $U_2$, respectively.  Explicitly, 
\begin{eqnarray}
U_1^{\dagger}(t_1, 0) H_2 U_1(t_1, 0) = 
\frac{\hbar \omega}{2} \left( - \cos \phi \, \sigma_x + 
\sin \phi \, \sigma_y  \right).
\end{eqnarray}
This shows that both integrands on the right-hand side 
of Eq.~\eqref{eq:total_te} are 
time-independent, which implies 
\begin{eqnarray}
\delta_\pm & = & -\frac{\theta_1}{2} \bra{\psi_\pm} 
\sigma_y \ket{\psi_\pm} 
\nonumber \\ 
 & & - \frac{\theta_2}{2} 
\bra{\psi_\pm}  \left( - \cos{\phi} \, \sigma_x + 
\sin{\phi} \, \sigma_y  \right) \ket{\psi_\pm} = 
\pm \delta.
\end{eqnarray}
To find geometric gates, we need to solve for which choices 
of $\theta_1$, $\theta_2$, and 
$\phi$ the dynamical phases become trivial, i.e., satisfy 
the condition $2 \delta = 0, \! \! \mod 2 \pi$. We thus look for 
numerical solutions of 
\begin{eqnarray}
\theta_1 \bra{\psi_\pm} \sigma_y \ket{\psi_\pm} & + & 
\theta_2 \bra{\psi_\pm} \left( - \cos{\phi} \, \sigma_x 
+ \sin{\phi} \, \sigma_y  \right) \ket{\psi_\pm} 
\nonumber \\ 
 & = & 0, \! \! \mod 2 \pi.
\end{eqnarray}
We restrict to $2 \delta = 0$ in the following. 

It is possible to find any number of roots for the same set 
of $\{ \theta_1, \phi \}$. In Fig.~\ref{fig:Roots}, 
roots are shown for when $0 < \theta_1 < 2 \pi$ and $0 < \theta_2 < 3 \pi$ as 
functions of $\phi$. A positive $\theta_1$ ($\theta_2$) corresponds 
to a clockwise rotation driven by the first (second) pulse. 
Roots can be found for counter-clockwise rotations 
to have the same shape but reflected or inverted. For $\theta_1 > 0$,  
$\theta_2 < 0$, the roots are inverted through the point 
$(\phi,\theta_2) = (\pi, 0)$; for $\theta_1 < 0$,  
$\theta_2 > 0$, on the other hand, they are reflected
in the line $\phi = \pi$. When both rotations are counter-clockwise the roots 
are reflected in the line $\theta_1 = 0$. In Fig.~\ref{fig:GeometricPhase}, 
the GPs corresponding to these roots are shown. For both 
counter-clockwise rotations where $\theta_1 > 0$, 
$\theta_2 < 0$ and $\theta_1 < 0$,  $\theta_2 > 0$ the 
GPs are reflected in the line $\phi = \pi$ and for the case when both rotations 
are counter-clockwise the GPs are the same as in Fig.~\ref{fig:GeometricPhase}. 

In the upper panel of Fig.~\ref{fig:Roots}, 
one can see that there are three different choices of $\theta_2$ that 
give trivial dynamical phase for $\theta_1 = 0.9 \pi$. They can be found in narrow 
$\phi$-intervals near $\phi \sim 1.15 \pi$ and $\phi \sim 1.85 \pi$, in between the 
two sets of dashed lines in the figure. As can be seen in the upper panel of 
Fig.~\ref{fig:GeometricPhase}, these choices correspond to different 
GPs and would thus result in different GNGGs. Furthermore, the width of the $\phi$ intervals 
increases when $\theta_1$ approaches $\pi$, until the curve splits 
into two for $\theta_1 = \pi$, where the first rotation flips the computational 
states $\ket{0}$ and $\ket{1}$: one concave curve with minimum at 
$(\phi,\theta_2)=(1.5 \pi,\pi)$ (truncated at $\theta_2=3\pi$) and one horizontal 
line at $\theta_2=\pi$ that  corresponds to a `orange slice' shaped loop 
\cite{tian04,thomas11}. The latter is the special case where the computational 
basis and the eigenbasis coincide. In this cyclic case, the GP will be a linear 
function of the opening angle $\phi$ \cite{remark3}, which is indeed visible as 
straight lines in Fig.~\ref{fig:GeometricPhase}. 

One motivating aspect of noncyclic geometric schemes is 
that they may shortened the exposure to decoherence effects 
by shortening the run time of the gates. To test this in 
the two-pulse realization of GNGGs, we use the total 
precession angle $\theta_1 + \theta_2$ as a natural 
measure of run time. A closer inspection of 
Fig.~\ref{fig:Roots} entails that this angle is at 
least $2 \pi$ for all roots. It thus appears that  
the run time cannot be shortened in the two-pulse noncyclic 
scheme, as compared to cyclic geometric gates with  
the pulse pair driving the computational basis states along 
`orange slice' shaped loops on the Bloch sphere 
\cite{tian04,thomas11}.

\begin{figure}[h]
\centering
\includegraphics[width = 0.3\textwidth]{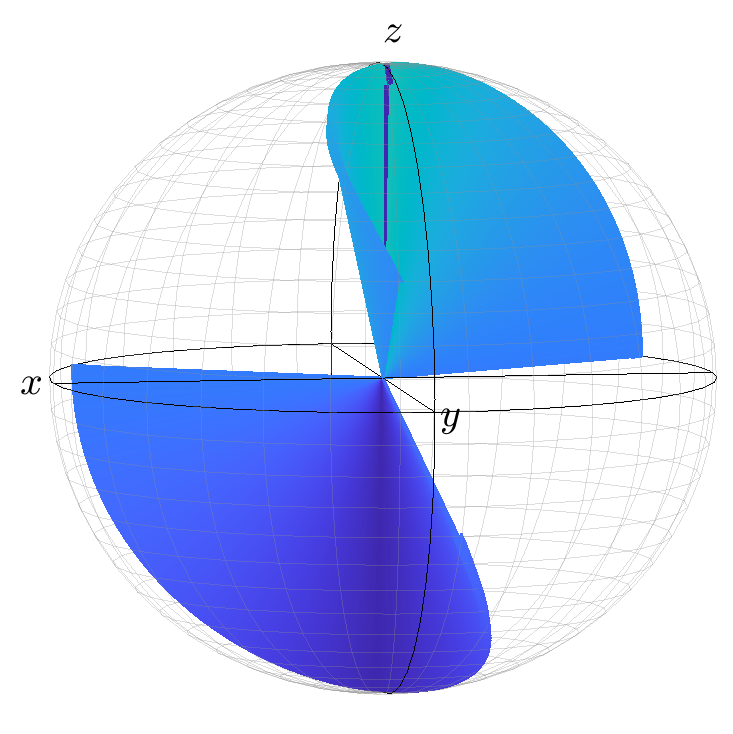}
\caption{Continuous set of eigenvectors of gates with $2 \gamma = \pi$ with 
the first rotation taken around the $y$-axis. The eigenvectors are swept so as 
to form surfaces inside the Bloch sphere of the qubit.}
\label{fig:PiRotations}
\end{figure}

\begin{figure}[h!]
\centering
\includegraphics[width=0.3\textwidth]{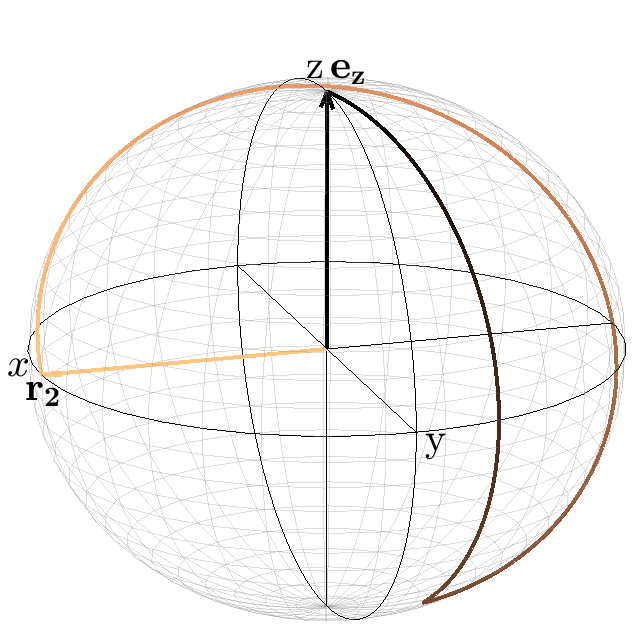}
\includegraphics[width=0.3\textwidth]{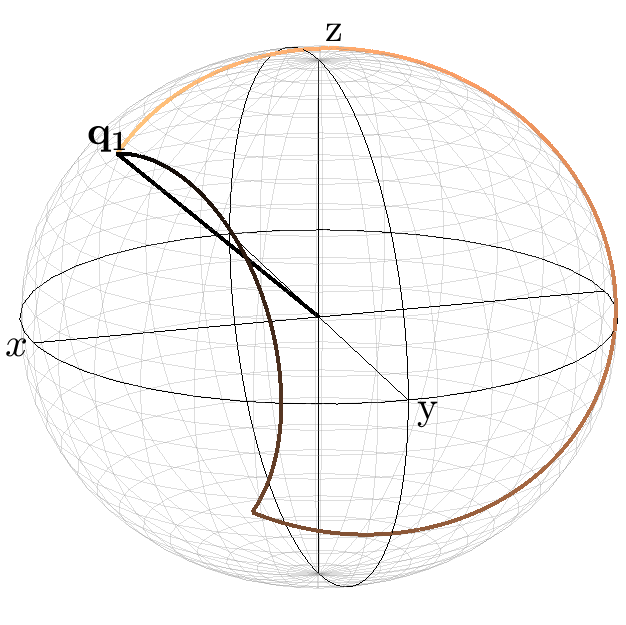}
\caption{The path of the $\ket{0}$-state (upper panel) and one of the eigenstates 
(lower panel) when acted upon by a geometric Hadamard gate. The paths gradually 
shifts from black (dark) at the start to yellow (bright) at the end. The computational 
state evolves along geodesic segments.} 
\label{fig:rotatedHadamard}
\end{figure}

To implement a specific gate, both eigenvalues and eigenvectors 
must match the desired gate. We have shown that it is possible 
to find any eigenvalue but it is also necessary to find the 
corresponding eigenvectors. To demonstrate this, eigenvectors 
corresponding to $2 \gamma = \pi$ are displayed in 
Fig.~\ref{fig:PiRotations}. These 
eigenvectors cover the entire $z$-axis. Keep in mind that 
only a first rotation around the $y$-axis has been 
considered so far. GNGGs with the same 
eigenvalue and an eigenvector only differing by a rotation around the $z$-axis 
can be found by changing the first axis of rotation 
to another in the $xy$-plane, while keeping the relation 
between the first and the second axis. With this finding 
any GNGGs with eigenvalue $2 \gamma = \pi$, 
such as the Hadamard gate $\mathbb{H}$, can be realized 
with only two pulses and it can similarly be shown for 
other $\gamma$ as well.

To give an example, we show how a Hadamard gate can be 
implemented. For this, $2 \gamma = \pi$ and eigenvectors correspond to 
$\pm \frac{{\bf e}_x+{\bf e}_z}{\sqrt{2}}$ on the Bloch sphere. 
These eigenvectors are not present in 
Fig.~\ref{fig:PiRotations}, but can be found by a suitable 
rotation around the $z$ axis. Fig.~\ref{fig:rotatedHadamard} 
illustrates this Hadamard gate acting on the computational 
state $\ket{0}$ and one of the eigenstates.

\section{Conclusions}
We have proposed a notion of noncyclic geometric gates 
in which both the computational basis and the eigenbasis 
acquire purely geometric phases. These two conditions require as 
precise control of pulse lengths as in the case of cyclic geometric gates, 
where cyclic evolution of the computational states needs to be ensured. 
We have demonstrated a physical realization of such {\it genuinely noncyclic 
geometric gates} in the single qubit case. This is 
achieved by using pulse pairs that drive 
the computational states along pairs of geodesic segments 
on the Bloch sphere, and simultaneously make the 
dynamical phase difference of the eigenstates to vanish. 
The proposed concept removes the ambiguity of standard 
noncyclic geometric gates 
\cite{friedenauer03,ericsson08,wang09,liu20,ji21,qiu21,sun22,guo22}, 
in which the computational basis undergoes purely geometric 
evolution, while the eigenstates generally do not. 
Our scheme takes advantage of noncyclic geometric phase in 
order to achieve universality. 

While the analysis focuses on gates using only two pulses,  
it can straightforwardly be extended to three or more pulses. 
This may open up for reduction of accumulated rotation angle 
as well as more elaborate paths of the computational states, 
so as to reduce the detrimental effect of noise and decoherence. 
The scheme may further be extended to nongeodesic evolution 
of the computational basis to further improve the error resilience 
of the gates. This extension would require simultaneous removal 
of the dynamical phase effects of the computational basis and 
the eigenstates of the gates. While we in this work have focused on 
the single-qubit case, the genuinely noncyclic geometric scheme can 
be extended to more general gates involving many qubits or even qudits.  

\section*{Acknowledgment}
E.S. acknowledges financial support from the Swedish Research Council 
(VR) through Grant No. 2017-03832.


\begin{thebibliography}{99} 
\bibitem{debnath16} S. Debnath, N. M. Linke, C. Figgatt, 
K. A. Landsman, K. Wright, and C. Monroe, 
Demonstration of a small programmable quantum computer with 
atomic qubits, 
Nature (London) {\bf 536}, 63 (2016).  
\bibitem{watson18} T. F. Watson, S. G. J. Philips, E. Kawakami, 
D. R. Ward, P. Scarlino, M. Veldhorst, D. E. Savage, M. G. Lagally, 
M. Friesen, S. N. Coppersmith, M. A. Eriksson, and 
L. M. K. Vandersypen, 
A programmable two-qubit quantum processor in silicon, 
Nature (London) {\bf 555}, 633 (2018).   
\bibitem{wu19} Y. Wu, Y. Wang, X. Qin, X. Rong, and J. Du, 
A programmable two-qubit solid-state quantum processor under 
ambient conditions,  
npj Quantum Inf. {\bf 5}, 9 (2019). 
\bibitem{roy20} T. Roy, S, Hazra, S. Kundu, M. Chand, M. P. Patankar, 
and R. Vijay, 
Programmable Superconducting Processor with Native Three-Qubit Gates, 
Phys. Rev. Appl. {\bf 14}, 014072 (2020). 
\bibitem{nielsen00} M. A. Nielsen and I. L. Chuang, 
{\it Quantum Computation and Quantum Information} 
(Cambridge University Press, Cambridge, UK, 2000), Ch. 10.6.
\bibitem{gottesman10} D. Gottesman, 
An introduction to quantum error correction and fault-tolerant quantum computation, 
Proc. Symp. Appl. Math. {\bf 68}, 13 (2010). 
\bibitem{jones00} J. A. Jones, V. Vedral, A. Ekert, and G. Castagnoli, 
Geometric quantum computation using nuclear magnetic resonance, 
Nature (London) {\bf 403}, 869 (2000).
\bibitem{ekert00} A. Ekert, M. Ericsson, P. Hayden, H. Inamori, 
J. A. Jones, D. K. L. Oi, and V. Vedral, 
Geometric quantum computation, 
J. Mod. Opt. {\bf 47}, 2501 (2000).
\bibitem{xiang01} W. Xiang-Bin and M. Keiji, 
Nonadiabatic Conditional Geometric Phase Shift with NMR, 
Phys. Rev. Lett. {\bf 87}, 097901 (2001). 
\bibitem{zhu02} S.-L. Zhu and Z. D. Wang, 
Implementation of Universal Quantum Gates Based on Nonadiabatic Geometric Phases,
Phys. Rev. Lett. {\bf 89}, 097902 (2002).
\bibitem{zhu03} S.-L. Zhu and Z. D. Wang, 
Unconventional Geometric Quantum Computation, 
Phys. Rev. Lett. {\bf 91}, 187902 (2003). 
\bibitem{wang07} Z. S. Wang, C. Wu, X.-L. Feng, L. C. Kwek, C. H. Lai, C. H. Oh, and V. Vedral, 
Nonadiabatic geometric quantum computation, 
Phys. Rev. A {\bf 76}, 044303, (2007). 
\bibitem{liu19} B.-J. Liu, X.-K. Song, Z.-Y. Xue, X. Wang, and M.-H. Yung, 
Plug-and-Play Approach to Non-adiabatic Geometric Quantum Gates, 
Phys. Rev. Lett. {\bf 123}, 100501 (2019).
\bibitem{berry84} M. V. Berry,
Quantal phase factors accompanying adiabatic changes, 
Proc. R. Soc. London Ser. A {\bf 392}, 45 (1984).
\bibitem{aharonov87} Y. Aharonov and J. Anandan, 
Phase change during a cyclic quantum evolution, 
Phys. Rev. Lett. {\bf 58}, 1593 (1987). 
\bibitem{leibfried03} D. Leibfried, B. DeMarco, V. Meyer, D. Lucas, M. Barrett, 
J. Britton, W. M. Itano, B. Jelenkovi\'c, C. Langer, T. Rosenband, and D. J. Wineland, 
Experimental demonstration of a robust, high-fidelity geometric two ion-qubit phase gate, 
Nature {\bf 422}, 412 (2003).
\bibitem{zhu05} S.-L. Zhu and P. Zanardi, 
Geometric quantum gates that are robust against stochastic 
control errors, 
Phys. Rev. A {\bf 72}, 020301(R) (2005).
\bibitem{ota09} Y. Ota and Y. Kondo, 
Composite pulses in NMR as nonadiabatic geometric quantum gates, 
Phys. Rev. A {\bf 80}, 024302 (2009). 
\bibitem{kondo11} Y. Kondo and M. Bando, Geometric quantum gates, composite pulses, 
and Trotter-Suzuki formulas, J. Phys. Soc. Jpn. {\bf 80},  054002 (2011). 
\bibitem{ichikawa12} T. Ichikawa , M. Bando , Y. Kondo, and M. Nakahara, 
Geometric aspects of composite pulses, 
Phil. Trans. R. Soc. A {\bf 370}, 4671 (2012). 
\bibitem{berger13} S. Berger, M. Pechal, A. A. Abdumalikov, Jr., 
C. Eichler, L. Steffen, A. Fedorov, A. Wallraff, and S. Filipp, 
Exploring the effect of noise on the Berry phase, 
Phys. Rev. A {\bf 87}, 060303(R) (2013).
\bibitem{ding22} C.-Y. Ding, L.-N. Ji, T. Chen, and Z.-Y. Xue, 
Path-optimized nonadiabatic geometric quantum computation on superconducting qubits, 
Quantum Sci. Technol. {\bf 7}, 015012 (2022). 
\bibitem{liang22} M.-J. Liang and Z.-Y. Xue, 
Robust nonadiabatic geometric quantum computation by dynamical correction, 
Phys. Rev. A {\bf 106}, 012603 (2022).
\bibitem{friedenauer03} A. Friedenauer and E. Sj\"oqvist, 
Noncyclic geometric quantum computation, 
Phys. Rev. A {\bf 67}, 024303 (2003).
\bibitem{ericsson08} M. Ericsson, D. Kult, E. Sj\"oqvist, and 
J. {\AA}berg, 
Nodal free geometric phases: Concept and application to geometric 
quantum computation, 
Phys. Lett. A {\bf 372}, 596 (2008). 
\bibitem{wang09} Z. S. Wang, G. Q. Liu, and Y. H. Ji, 
Noncyclic geometric quantum computation in a 
nuclear-magnetic-resonance system, 
Phys. Rev. A {\bf 79}, 054301 (2009). 
\bibitem{liu20} B. Liu, S. Su, and M. Yung, 
Nonadiabatic noncyclic geometric quantum computation in 
Rydberg atoms, 
Phys. Rev. Res. {\bf 2}, 043130 (2020).
\bibitem{ji21} L.-N. Ji, C.-Y. Ding, T. Chen, and Z.-Y. Xue
Noncyclic geometric quantum gates with smooth paths via 
invariant-based shortcuts, 
Adv. Quantum Tech. {\bf 2}, 2100019 (2021). 
\bibitem{qiu21} L. Qiu, H. Li, Z. Han, W. Zheng, X. Yang, Y. Dong, 
S. Song, D. Lana, X. Tana, and Y. Yu, 
Experimental realization of noncyclic geometric gates with 
shortcut to adiabaticity in a superconducting circuit, 
Appl. Phys. Lett. {\bf 118}, 254002 (2021).
\bibitem{sun22} L. N. Sun, F. Q. Guo, Z. Shan, M. Feng, 
L. L. Yan, and S. L. Su
One-step implementation of Rydberg nonadiabatic noncyclic geometric 
quantum computation in decoherence-free subspaces, 
Phys. Rev. A {\bf 105}, 062602 (2022). 
\bibitem{guo22} F.-Q. Guo, X.-Y. Zhu, M.-R. Yun, L.-L. Yan, 
Y. Zhang, Y. Jia1,3 and S.-L. Su, 
Multiple-qubit nonadiabatic noncyclic geometric quantum computation 
in Rydberg atoms, 
EPL {\bf 137}, 55001 (2022). 
\bibitem{zhang21} J. W. Zhang, L.-L. Yan, J. C. Li, G. Y. Ding, 
J. T. Bu, L. Chen, S.-L. Su, F. Zhou, and M. Feng, 
Single-Atom Verification of the Noise-Resilient and Fast 
Characteristics of Universal, Nonadiabatic Noncyclic Geometric 
Quantum Gates, 
Phys. Rev. Lett. {\bf 127}, 030502 (2021). 
\bibitem{samuel88} J. Samuel and R. Bhandari, 
General Setting for Berry's Phase, 
Phys. Rev. Lett. {\bf 60}, 2339 (1988). 
\bibitem{mukunda93} N. Mukunda and R. Simon, 
Quantum kinematic approach to the geometric phase. I. 
General formalism, 
Ann. Phys. (N.Y.) {\bf 228}, 205 (1993). 
\bibitem{remark1} Note that the chosen phases of the vectors 
$\ket{A} = \frac{1}{\sqrt{2}} \big( \ket{1} - \ket{0} \big)$ 
and $\ket{B} = \ket{1}$ in the second term are made such that 
their Pancharatnam relative phase $\arg \langle A\ket{B}$ 
\cite{pancharatnam56} vanishes. This is to assure that $e^{i\pi}$ 
is the GP factor given the assumption that $\mathbb{H}$ is a GNGG. 
\bibitem{pancharatnam56} S. Pancharatnam, 
Generalized theory of interference, and its applications. 
Part I. Coherent pencils, 
Proc. Indian Acad. Sci., Sect. A {\bf 44}, 247 (1956). 
\bibitem{sakurai93} J. J. Sakurai, 
{\it Modern Quantum Mechanics}, 2nd Ed.
(Addison Wesley Longman, Reading, MA, 1993).  
\bibitem{remark2} To see this, we note that the eigenstates and 
computational states coincide for cyclic geometric gates, thus 
resulting in an Abelian subset consisting of phase shift gates $\ket{q} \mapsto 
e^{i\gamma_q} \ket{q}$, $q=0,1$. 
\bibitem{tian04} M. Tian, Z. W. Barber, J. A. Fischer, and 
W. R. Babbitt, 
Geometric manipulation of the quantum states of two-level atoms, 
Phys. Rev. A {\bf 69}, 050301(R) (2004).
\bibitem{thomas11} J. T. Thomas, M. Lababidi, and M. Z. Tian, 
Robustness of single-qubit geometric gate against systematic error, 
Phys. Rev. A {\bf 84}, 042335 (2011).
\bibitem{remark3} To see this, we note that an `orange slice' loop with opening angle 
$\phi$ encloses the solid angle $\Omega = 2\phi$, which implies the cyclic GPs 
$\Gamma_0 = -\Gamma_1 = -\frac{1}{2} \Omega = -\phi$. 
\end{thebibliography}
\end{document}